\documentclass[12pt,amsbsy]{article}
\usepackage{graphicx}
\newcommand{\beq}{\begin{equation}}
\newcommand{\eeq}{\end{equation}}
\newcommand{\beqn}{\begin{eqnarray}}
\newcommand{\eeqn}{\end{eqnarray}}

\newcommand{\lsim}{\mbox{$<$\hspace{-0.8em}\raisebox{-0.4em}{$\sim$}}}

\newcommand{\de}{\mbox{${\delta}$}}

\begin{document}

\begin{titlepage}

\vspace{1cm}

\begin{center}
{\bf \large IS RADIATION OF QUANTIZED BLACK HOLES OBSERVABLE?}
\end{center}

\vspace{5mm}

\begin{center}
I.B. Khriplovich \footnote{khriplovich@inp.nsk.su}\\
{\em Budker Institute of Nuclear Physics\\
11 Lavrentjev pr., 630090 Novosibirsk, Russia,\\
and Novosibirsk University}
\end{center}

\begin{center}
N. Produit \footnote{Nicolas.Produit@obs.unige.ch}\\
{\em INTEGRAL Science Data Center\\
16, Chemin d'Ecogia, CH-1290 Versoix, Switzerland}
\end{center}

\bigskip

\begin{abstract}
If primordial black holes (PBH) saturate the present upper limit
on the dark matter density in our Solar system and if their
radiation spectrum is discrete, the sensitivity of modern
detectors is close to that necessary for detecting this radiation.
This conclusion is not in conflict with the upper limits on the
PBH evaporation rate.

\end{abstract}


\end{titlepage}

{\bf 1.} We discuss in the present note the possibility to detect
the radiation of primordial black holes (PBH), under the
assumption that they constitute a considerable part of dark
matter. We consider the situation when this radiation has a
discrete spectrum, as predicted in some models of quantized black
holes.

The idea of quantizing the horizon area of black holes was put
forward by Bekenstein in the pioneering article~\cite{bek}. The
idea is quite popular now, but the list of references on the
subject is too long to be presented in this note.

We will essentially rely here on the results of~\cite{khr}. It was
demonstrated therein, under quite natural general assumptions,
that the spectrum of black hole radiation is discrete and fits the
Wien profile (see also in this relation earlier
article~\cite{bem}), and that the natural widths of the lines are
much smaller than the distances between them. This spectrum starts
with a line of a typical frequency close to the Hawking
temperature $T=1/(8\pi k m)$ ($k$ is the Newton gravitational
constant, $m$ is the black hole mass), and due to the exponential
fall-down, the spectrum consists effectively of 2 -- 3 lines only,
separated by intervals also close to~$T$. However, the total
intensity of these few lines, situated around the maximum of the
Planck profile, is about the same as that of the Hawking thermal
radiation, the latter being saturated essentially just by this
region.

\vspace{5mm}

{\bf 2.} The analysis of the observational data for the secular
perihelion precession of Earth and Mars results in the following
upper limit on the density $\rho_{\rm{dm}}$ of dark matter in the
Solar system~\cite{kp}:
\begin{equation}\label{re}
\rho_{\rm{dm}}^{\rm{ss}} < 3 \times 10^{-19}\; {\rm g/cm^3}\,.
\end{equation}
This limit is based on the precision EPM ephemerides constructed
in~\cite{pit1}, and on the possible deviations~\cite{pit2} of the
results of theoretical calculations from the observational data
for the planets obtained from about 250000 high-precision American
and Russian ranging to planets and spacecraft. Of course, limit
(\ref{re}) is quite modest as compared to the galactic dark matter
density $\rho_{\rm{dm}}^{\rm{\,g}} \simeq 0.5 \times 10^{-24}\;
{\rm g/cm^3}$, to say nothing of the cosmological one
$\rho_{\rm{dm}}^{\rm{\,c}} \simeq 0.4 \times 10^{-29}\; {\rm
g/cm^3}$. However, in the absence of better observational data on
$\rho_{\rm{dm}}^{\rm{ss}}$, we will rely below on limit (\ref{re})
(having in mind, in particular, the huge difference of scales
between $\rho_{\rm{dm}}^{\rm{\,g}}$
and~$\rho_{\rm{dm}}^{\rm{\,c}}$).

Our quantitative estimates for the expected signal from PBHs are
performed under the optimistic assumption that their density $\rho
\sim 10^{-19}\; {\rm g/cm^3}$. The results of these estimates are
presented in Table~1.
\begin{table}
\begin{center}
\begin{tabular}{|c|c|c|c|c|c|} \hline
 & & & & &  \\
\hspace{2mm} $m$, g \hspace{2mm} & \hspace{1mm} $n$, cm$^{-3}$
\hspace{1mm} &\hspace{3mm} $\bar{r}$, cm\hspace{3mm}&\hspace{1mm}
$T$, MeV\hspace{1mm} &\hspace{1mm} $N$,  ph
s$^{-1}$\hspace{3mm}&\hspace{1mm} $\nu$,  ph cm$^{-2}$
s$^{-1}$ \hspace{1mm}\\
 & & & & &  \\ \hline
 & & & & &  \\
$5\times 10^{14}$ & $2\times 10^{-34}$ & $1.7 \times 10^{11}$ & 20
& $6 \times 10^{19}$ & $1.6\times 10^{-4}$ \\
 & & & & &  \\ \hline
 & & & & &  \\
$2\times 10^{15}$ & $5\times 10^{-35}$  & $2.7\times 10^{11}$ & 5
 &  $1.5 \times 10^{19}$ & $1.6 \times 10^{-5}$\\
 & & & & &  \\ \hline
 & & & & &  \\
$10^{16}$  & $10^{-35}$ & $0.5 \times 10^{12}$  & 1  &  $3 \times
10^{18}$  &  $10^{-6}$\\
 & & & & &  \\ \hline
 & & & & &  \\
$10^{17}$  &$10^{-36}$  & $10^{12}$  &  0.1 &
$3 \times 10^{17}$  &  $2 \times 10^{-8}$ \\
 & & & & &  \\ \hline
\end{tabular}
\caption{Predictions for radiation of primordial black holes in
Solar system}
\end{center}
\end{table}
We confine here to four values for the black hole mass $m$,
starting with $5 \times 10^{14}$ g. It is well-known that a PBH
with the initial mass $\;m_0 \;\lsim \;5 \times 10^{14}$ g just
could not survive till our time due to the same radiation. On the
other hand, for masses essentially larger than $10^{17}$ g, the
signal gets hopelessly small. In Table 1, $n= \rho/m$ is the
density of number of primordial black holes, $\bar{r} = n^{-1/3}$
is the typical distance between two of them;
\beq\label{N}
N \simeq \,\frac{1}{64 \pi^4}\,\frac{c}{r_g}
\eeq
is the number of quanta per second emitted in a line by a black
hole with a gravitational radius $r_g$ (here we use the estimate
from~\cite{khr}, and retain for clarity the velocity of light
$c$); $\nu$ is the typical expected flux of quanta in a line at
the distance $\bar{r} = n^{-1/3}$. We have mentioned already that
the temperature $T$ roughly corresponds to the energy of the first
line in the discrete spectrum of a black hole. With 2--- 3
relatively strong lines in a spectrum, one may expect gamma lines
with energies about 20, 40, 60 MeV in the spectra of PBHs with
mass $5 \times 10^{14}$~g; about 5, 10, 15 MeV in the spectra of
PBHs with mass $2 \times 10^{15}$ g; about 1, 2, 3 MeV in the
spectra of PBHs with mass $10^{16}$ g; about 100, 200, 300 KeV in
the spectra of PBHs with mass $10^{16}$ g. The presence of 2~---~3
roughly equidistant narrow lines in the spectrum of a source of
radiation is essential for its identification with a quantized
black hole.

Experimentally, to look for such a source we need a gamma ray
telescope that is able to do fine spectroscopy for point sources.
One candidate here could be the IBIS imager on board INTEGRAL, its
point spread function being 12 arcmin. With the typical distance
to the nearest source 10$^{12}$~cm and typical speed of object in
the solar system 10$^6$ cm s$^{-1}$, such a source will displace
itself relative to the observing satellite at an angular velocity
of 10$^{-6}$~rad~s$^{-1}$ in an unpredictable direction. This
limits the useful integration time of the search for such a source
to 10$^4$ s. Degrading the angular resolution of the images would
allow to stack the images longer, but will decrease the contrast
by increasing the noise, so that the limit will not improve. Using
the last official IBIS sensitivity data~\cite{ao5}, and
recalculating them for a 10$^4$ s exposure, we obtain the
sensitivity curve of Figure 1. The best sensitivity of IBIS for an
unresolved line of a point source amounts to $1.7 \times
10^{-\,4}$ ph cm$^{-2}$ s$^{-1}$ at 100 KeV. Clearly, the IBIS
sensitivity is insufficient for our purpose, by about 4 orders of
magnitude at 100~KeV, and by about 3 orders of magnitude at 1 Mev.
We note that IBIS never detected any convincing point source with
discrete lines.

Another possibility is to use the SPI spectrometer on board
INTEGRAL, which has better sensitivity for discrete
spectrum~\cite{spi}. Its imaging resolution being worse, one can
stack here images up to 10$^{6}$ seconds without loss of
sensitivity. The sensitivity curve of SPI varies wildly with
energy due to the presence of numerous gamma ray emission lines in
the background spectrum, but stays always within the shaded area
in Figure 1. Thus, for all energies SPI is more sensitive than
IBIS. In particular, the SPI sensitivity is sufficient to observe
the line around 5 MeV which belongs to a PBH with $m \sim 2 \times
10^{15}$ g. This would require to analyze very deep sky images in
a large set of fine energy bands, which is time consuming, but
doable in some sky direction with existing data. If a signal will
be found, still one should exclude the existence of an unexpected
background feature.

One may wonder also about the quantized radiation of galactic
primordial black holes. However, the typical distances to PBHs of
the Galaxy are much larger than distances to PBHs of the Solar
system. Then, even the galactic dark matter density
$\rho_{\rm{dm}}^{\rm{\,g}} \simeq 0.5 \times 10^{-24}\; {\rm
g/cm^3}$, to say nothing of its possible PBH component, is
certainly much lower than $\rho \sim 10^{-19}\; {\rm g/cm^3}$
assumed above. Therefore, one can measure here only a collective
effect of many PBHs, but not the radiation of a single PBH. In
this case, it does not look realistic at all to resolve the
discrete spectrum of a single black hole.

\begin{figure}[t]
\centering
\includegraphics[width=0.8\linewidth]{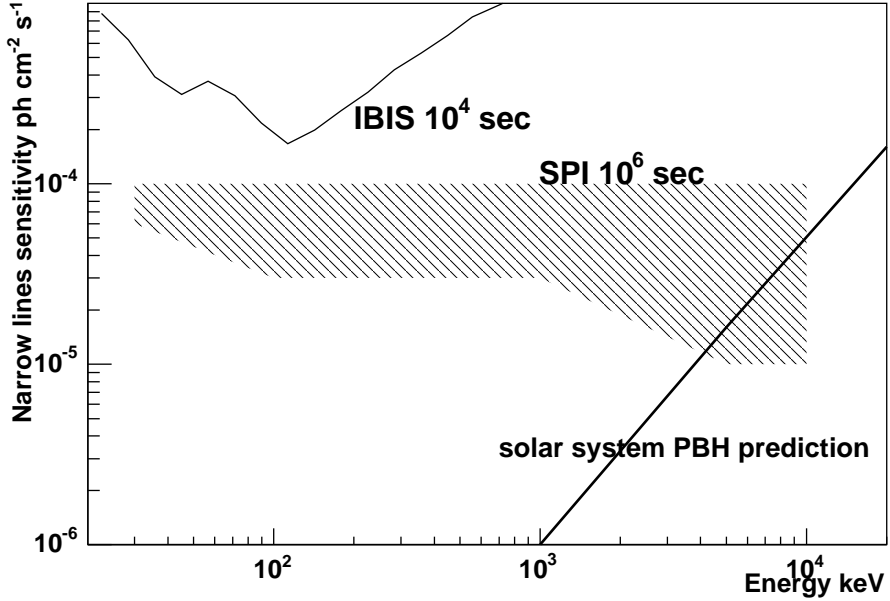}
\caption{Sensitivity of IBIS imager for narrow line of point-like
source exposed for 10$^4$ s and of the SPI spectrometer for point-like
source exposed for 10$^6$ s}
\end{figure}

\vspace{5mm}

{\bf 3.} Let us discuss now whether our assumption
\beq\label{as}
\rho \sim 10^{-19}\; {\rm g/cm^3}
\eeq
for the density of primordial black holes in the Solar system is
compatible with other searches for PBHs. The direct searches for
the bursts of gamma rays expected from their evaporation result in
upper limits on the PBH evaporations\footnote{According to
\cite{lin}, these limits are the most model--independent ones.} on
the level \cite{al} -- \cite{lin}
\beq\label{ex}
10^6 \;{\rm pc}^{-3} \; {\rm yr}^{-1}.
\eeq
These results constrain directly only the number density of very
light PBHs with typical life time about one year, which was the
typical observation time. To relate it to the number density of
much heavier (but still light) black holes of interest to us, we
need to know the PBH mass distribution. For the estimates we
assume, following~\cite{carr}, that it is as follows:
\beq\label{dis1}
d n = \de \rho_{\rm{dm}}\,\frac{1}{2}\,m_0^{1/2} m_1^{-5/2}
dm_1\,,\quad m_1
> m_0\,.
\eeq
Here the factor $\de$ is the relative contribution of PBHs to the
dark matter (we put above effectively $\de \sim 1$). Distribution
(\ref{dis1}) is normalized in such a way that, being integrated
with the weight $m_1$, it gives $\rho = \de \rho_{\rm{dm}}$, the
mass density of PBHs. We have labelled the mass $m$ in this
distribution with index 1 to demonstrate that it refers in fact
not to the present mass distribution, we are interested in, but to
the mass distribution after the formation of PBHs. The cut-off at
$m_1 = m_0 = 0.5 \times 10^{15}\;{\rm g}$ reflects the mentioned
fact that a PBH with a smaller mass will not survive till present.
The behavior of distribution (\ref{dis1}) at the masses less than
$m_0$, including the effective cut-off of the divergence at small
$m_1$, is not directly related to our problem.

The initial mass $m_1$ of a black hole is related to its
contemporary one $m$ by relation
\beq\label{rel}
m_1^3 = m^3 + m_0^3\,;
\eeq
it follows immediately from the differential equation $\;\;dm/dt
\sim - 1/m^2\;$ that describes the evolution of the mass of a
black hole due to its radiation. When rewritten in terms of $m$,
distribution (\ref{dis1}) transforms into
\beq\label{dis}
d n = \de \rho_{\rm{dm}}\,\frac{1}{2}\,m_0^{1/2} m^2 (m^3 +
m_0^3)^{-3/2} dm\,,\quad m > 0\,.
\eeq
Though this mass spectrum extends formally to $m = 0$, it
decreases rapidly for small masses. Still, its maximum is at $m =
(4/5)^{1/3} m_0 \simeq 0.9\, m_0$, i.e. lies somewhat below $m_0$.

Let us estimate now the number density of PBHs exploding during a
year under the assumption of mass distribution (\ref{dis}). The
radiative life-time of a black hole with mass $m$ is well known to
be proportional to $m^3$. Since for $m = m_0$ the life-time is
$\;\sim 10^{10}$ years, for the mass $\mu$ of the very light PBH
of interest, we obtain
\[
\left(\frac{\mu}{m_0}\right)^3 \sim 10^{-10}\,.
\]
Then, with distribution (\ref{dis}), the number density of
these light PBHs is at present
\beq\label{dl}
n_{\mu} = \de \rho_{\rm{dm}} \,\frac{1}{2}\,m_0^{1/2} \int_0^{\mu}
d m\,m^2 (m^3 + m_0^3)^{-3/2} = \,\frac{1}{6}\,\de\,\frac{
\rho_{\rm{dm}}}{m_0}\,\left(\frac{\mu}{m_0}\right)^3 \sim
10^{-11}\,\de\,\frac{\rho_{\rm{dm}}}{m_0}\,.
\eeq
With $\rho_{\rm{dm}} \sim 10^{-19}\; {\rm g/cm^3}$ and $m_0 = 0.5
\times 10^{15}\;{\rm g}$, to comply with the upper limit
(\ref{ex}), we should put here $\de \sim 2 \times 10^{-5}$,
instead of our assumption $\de \sim 1$. However that upper limit
(\ref{ex}) corresponds to typical distances between evaporating
black holes
\beq\label{r1}
\bar{r}_1 \sim 10^{-2}\;{\rm pc} \sim 2 \times 10^3\;{\rm au}\,.
\eeq
But such distances are too large. There are no serious reasons to
expect that our initial assumption $\;\rho~\sim~10^{-19}$~g/cm$^3$
for the PBH density, being valid for the distances about 1~au from
the Sun, should be true for much larger distances $\sim 2 \times
10^3\;{\rm au}$. It looks quite natural that at such large
distances the PBH density is lower by orders of magnitude than
$\;\rho~\sim~10^{-19}$~g/cm$^3$.

On the other hand, even with our assumption
$\;\rho~\sim~10^{-19}$~g/cm$^3$ for the PBH density, we arrive
with equation (\ref{dl}) at typical distances between evaporating
black holes
\beq\label{r2}
\bar{r}_2 \sim 10^2\;{\rm au}\,,
\eeq
which is still much larger than the distances about 1 au from the
Sun, we are interested in.

Therefore, the observational results of \cite{al} -- \cite{lin} do
not exclude the possibility of existence of the point-like sources
of radiation with discrete spectrum, i.e. of quantized PBHs, in
the Solar system.

\begin{center}***\end{center}
The investigation was supported in part by the Russian Foundation
for Basic Research through Grant No. 05-02-16627.

\end{document}